\documentclass[12pt]{article}
\usepackage{amsfonts,amssymb,amsmath,graphicx,hyperref}

\newcommand{\EE}{\mathrm{e}} 
\newcommand{\II}{\mathrm{i}} 

\title{Comment on ``Time-like flows of energy-momentum and particle trajectories for the Klein--Gordon equation''}
\author{
	Roderich Tumulka\footnote{address: Mathematisches Institut 
	der Universit\"at M\"unchen, Theresienstr.\ 39, D-80333 M\"unchen, 
	Germany. E-mail: tumulka@mathematik.uni-muenchen.de}
	}

\begin{document}
\maketitle

\begin{abstract}
Horton, Dewdney, and Nesteruk \cite{Dewdney2} have proposed Bohm-type particle trajectories accompanying a Klein--Gordon wave function $\psi$ on Minkowski space. From two vector fields on space-time, $W^+$ and $W^-$, defined in terms of $\psi$, they intend to construct a timelike vector field $W$, the integral curves of which are the possible trajectories, by the following rule: at every space-time point, take either $W = W^+$ or $W = W^-$ depending on which is timelike. 

This procedure, however, is ill-defined as soon as both are timelike, or
both spacelike. Indeed, they cannot both be timelike, but they can well
both be spacelike, contrary to the central claim of \cite{Dewdney2}. We
point out the gap in their proof, provide a counterexample, and argue that,
even for a rather arbitrary wave function, the points where both $W^+$ and
$W^-$ are spacelike can form a set of positive measure.
\end{abstract}

Let $\psi=\EE^{P+\II S}$ (where $P$ and $S$ are real) solve the Klein-Gordon equation, $-\square \psi = m^2 \psi$. Set $P_\mu = \partial_\mu P$, $S_\mu = \partial_\mu S$, and 
$$
  \theta = \mbox{sinh}^{-1}\, \frac{P^\mu P_\mu - S^\mu S_\mu}{2P^\mu S_\mu}\,.
$$
That $P_\mu$ and $S_\mu$ are orthogonal is an exceptional case that we neglect, like the authors of \cite{Dewdney2}. For $W_\mu$ one is supposed to take either $W_\mu^+ = \EE^\theta P_\mu + S_\mu$ or $W_\mu^- = -\EE^{-\theta} P_\mu + S_\mu$, depending on which is timelike; they cannot both be timelike since they are orthogonal. The question is, could they both be spacelike? 

The authors of \cite{Dewdney2} declare that $W^+_\mu$ and $W^-_\mu$ cannot both be spacelike and argue like this: otherwise there exists a Lorentz frame such that $W^+_0 = 0$ and $W^-_0 = 0$, thus $\EE^\theta P_0 = -\EE^{-\theta} P_0$, from which they conclude $\EE^\theta = -\EE^{-\theta}$, which is impossible. 

It is correct that any two orthogonal spacelike vectors span a spacelike 2-plane (corresponding to $x^0 = 0$ in the appropriate Lorentz frame), but no contradiction arises since in this case $P_0$ would be $0$ (in this frame). This is the mistake in the proof. 

Together with $W^+_0 = 0$ (or $W^-_0 = 0$), $P_0 = 0$ implies $S_0 = 0$. Hence, for $W^+_\mu$ and $W^-_\mu$ to be spacelike it is necessary and sufficient that $P_\mu$ and $S_\mu$ span a spacelike 2-plane.

Can this case occur? Clearly: since
the Klein--Gordon equation is of second order, one may choose $\psi$ and
$\partial_0\psi$ \textit{ad libitum} on the $x^0 =0$ hyperplane. Can it
also occur for the first-order Klein--Gordon equation $-\II\partial_0 \psi
= \sqrt{m^2 -\Delta} \,\psi$, or, equivalently, for functions from the
positive-energy subspace? Here is an example: let $\psi$ be a superposition
of three\footnote{Two will not suffice for an example since $P_\mu$ and $S_\mu$ are linear combinations of the $k_\mu$ vectors.} plane waves 
$$
  \psi(x) = \sum_{i=1}^3 c_i \:\EE^{\II k_\mu^{(i)} x^\mu}
$$
with wave vectors $k_\mu^{(1)} = \Big(m,0,0,0\Big)$, $k_\mu^{(2)} =
\Big(\sqrt{27}m,\sqrt{26}m,0,0\Big)$,  $k_\mu^{(3)} =
\Big(\sqrt{27}m,0,\sqrt{26}m,0\Big)$, and $c_1 = 3$, $c_2 = -1/\sqrt{3} -
\II$, $c_3 = \II$. Then, at the coordinate origin, we find $P_\mu =
(0,\alpha,-\alpha,0)$ and $S_\mu = (0,-\beta,0,0)$ with $\alpha = \sqrt{26}
m/\gamma$, $\beta=\alpha/\sqrt{3}$ and $\gamma=3-1/\sqrt{3}$. This example
could also be made square-integrable by replacing the plane waves $\exp(\II
k_\mu^{(i)}x^\mu)$ by positive-energy $L^2$ Klein--Gordon functions $\varphi^{(i)}(x)$ with the properties $\varphi^{(i)}(0) = 1$ and $\partial_\mu \varphi^{(i)} (0) = \II k_\mu^{(i)}$.

One may suspect, however, that perhaps this particular wave function $\psi$
is very exceptional, and perhaps even that for this special wave function
the coordinate origin is a rather atypical point, so that the sort of
situation just described can be ignored. After all, we would be willing to ignore the case where $P_\mu$ and $S_\mu$ are orthogonal because in the 8-dimensional space of all possible pairs of vectors $P_\mu, S_\mu$ it corresponds to a subset of dimension 7, and therefore one would expect that the space-time points where this happens form a set of measure zero.

But since $W^+_\mu W^{+\mu}$ and $W^-_\mu W^{-\mu}$ are continuous
functions of $P_\mu$ and $S_\mu$, the set of pairs $P_\mu, S_\mu$ where
both $W^+$ and $W^-$ are spacelike is open (and nonempty) and thus has 
positive measure in 8 dimensions. I know of
nothing precluding any pairs $P_\mu,S_\mu$ from arising from a
Klein--Gordon wave function, so it seems reasonable to expect that the
space-time points with spacelike $W^+$ and $W^-$ form a set of positive
measure for many wave functions, perhaps for most.

\bigskip

\noindent \textit{Acknowledgements.} I wish to thank Sheldon Goldstein for improvements and simplifications of the arguments.

\end{document}